\documentclass[reprint,aps, pra,
 reprint,
 superscriptaddress,
 amsmath,
 amssymb,
 nobibnotes,
 longbibliography,
 floatfix,
]{revtex4-2}
\usepackage{fix-cm}
\usepackage{amsmath}
\usepackage{booktabs}
\usepackage{hyperref}
\hypersetup{hidelinks}
\usepackage{algorithm}
\usepackage{algpseudocode}
\usepackage{amssymb}
\usepackage{xcolor}
\usepackage[normalem]{ulem}
\usepackage{mathtools}
\usepackage{makecell}

\newcommand{\circumeq}{\mathrel{\widehat{=}}}

\usepackage{tikz}
\usetikzlibrary{positioning}
\usetikzlibrary{quantikz2}

\begin{document}

\title{Automated reduction of fault-tolerant circuits}

\author{Hyeongjun Jeon}
\affiliation{Department of Computer Science and Engineering, Seoul National University, Seoul 08826, Republic of Korea}
\affiliation{NextQuantum, Seoul National University, Seoul 08826, Republic of Korea}

\author{Jeonghun Lee}
\affiliation{Department of Computer Science and Engineering, Seoul National University, Seoul 08826, Republic of Korea}
\affiliation{NextQuantum, Seoul National University, Seoul 08826, Republic of Korea}

\author{Taehyun Kim}
\email{taehyun@snu.ac.kr}
\affiliation{Department of Computer Science and Engineering, Seoul National University, Seoul 08826, Republic of Korea}
\affiliation{NextQuantum, Seoul National University, Seoul 08826, Republic of Korea}
\affiliation{Institute of Applied Physics, Seoul National University, Seoul 08826, Republic of Korea}
\affiliation{Institute of Computer Technology, Seoul National University, Seoul 08826, Republic of Korea}
\affiliation{Automation and Systems Research Institute, Seoul National University, Seoul 08826, Republic of Korea}
\date{\today}

\begin{abstract}
We present an automated method for reducing fault-tolerant circuits through fault-equivalent rewrites. Starting from a known fault-tolerant circuit, the search applies non-reducing enabling rules to expose Bell-pair reductions, each of which removes an ancilla preparation and a \textsc{cnot} gate. Because every search transition preserves fault equivalence, the resulting circuits inherit the fault-tolerance properties of the input circuit. Candidate circuits are evaluated using circuit-level Monte Carlo simulation. For Shor-style syndrome extraction with the $[[7,1,3]]$ code, our method reduces one syndrome-measurement round from 30 to 18 ancilla preparations and from 54 to 42 \textsc{cnot} gates. At a physical two-qubit error rate of $p = 10^{-3}$, the optimized circuit lowers the logical error rate by approximately 21\% for both logical basis states. The reduction ranges from 13\% to 23\% over a range of $p$ spanning two orders of magnitude. We also apply the method to Steane-based dynamic syndrome
extraction constructed from Goto's verified logical-$\ket{0}$ preparation, in which the
verification qubit serves as a flag. The search selects a circuit
using four ancillas and 14 \textsc{cnot} gates, matching the resource counts of the previous design but
with an improved \textsc{cnot} depth. Under depolarizing idle noise at
$3p/10$ the selected circuit lowers the logical error rate by approximately $15\%$ for both logical
basis states, and by approximately $10\%$ at $p/10$. These results demonstrate that automated
fault-equivalent rewriting can identify circuits with lower resource costs or improved logical
performance without separately verifying the fault tolerance of every candidate.\end{abstract}
%\keywords{flag fault-tolerant circuits, automated reduction, quantum error correction, syndrome extraction}
\maketitle

\section{Introduction}

Fault-tolerant quantum computation requires encoded operations to be implemented so that a limited number of physical faults cannot cause an uncorrectable logical error \cite{aliferis2005quantum}. Establishing these conditions for a candidate circuit requires analyzing how faults at different circuit locations propagate to the output. The number of relevant fault configurations grows rapidly with the circuit size and the number of simultaneous faults considered \cite{chen2025verifying}. Consequently, fault-tolerant constructions are commonly organized into smaller components, or \emph{gadgets}, whose properties can be established independently and then composed \cite{aliferis2005quantum}. The need to check these properties for many candidate circuits also limits the scalability of search-based synthesis and optimization methods.

Several general constructions are known for implementing fault-tolerant gadgets for quantum error-correcting codes (QECCs), including Shor-style and Steane-style syndrome extraction \cite{shor1996fault,steane1997active}. Shor-style extraction uses cat states to measure stabilizers, whereas Steane-style extraction couples the data transversally to an encoded ancilla block. Although these constructions apply to broad classes of codes, their generality leaves considerable space for optimization when they are instantiated for a particular QECC. For example, ancilla-efficient constructions have been developed for syndrome extraction and cat-state preparation \cite{Prabhu2023faulttolerant}.

Automated optimization of concrete circuits subject to fault-tolerance constraints remains comparatively underdeveloped. Many low-overhead designs instead rely on carefully constructed circuit structures, such as flag qubits that detect faults capable of propagating to uncorrectable data errors
\cite{chao2018quantum,chamberland2018flag,chao2020flag,reichardt2021fault}.
Recent automated approaches have used reinforcement learning to discover fault-tolerant logical state preparation circuits \cite{zen2025quantum} and satisfiability-based optimization to synthesize depth- or gate-optimal preparation and verification circuits \cite{peham2025automated}. Satisfiability modulo theories has also been applied to the synthesis of fault-tolerant Clifford circuits \cite{shutty2022decoding}. In these approaches, fault tolerance must be incorporated into a reward function, encoded as a set of constraints, or checked for the resulting candidates. The associated cost can therefore grow with both the circuit search space and the set of relevant fault configurations. This motivates search methods in which fault tolerance is preserved by construction throughout the search.

Rodatz \textit{et al.} recently introduced fault equivalence and a collection of fault-equivalent ZX-calculus rewrites that preserve a circuit's behavior under undetectable faults \cite{rodatz2025fault}. This framework extends their earlier notion of distance-preserving rewrites \cite{rodatz2026floquetifying}. Po{\'o}r \textit{et al.} subsequently used these rewrites to derive a dynamic syndrome-extraction protocol for the $[[7,1,3]]$ Steane code \cite{poor2026ultra}. Their protocol uses two circuits: a fault-tolerant primary circuit consisting of 14 \textsc{cnot} gates and four ancillas, and a non-fault-tolerant recovery circuit consisting of 11 \textsc{cnot} gates and three ancillas; the \textsc{cnot} counts are optimal under the circuit-model assumptions stated in their work. These transformations are formulated using the ZX-calculus, a graphical language for representing and transforming quantum computations
\cite{coecke2008interacting,duncan2020graph}. Their circuit derivations rely on manually constructed rewrite sequences, and automating this process has been identified as necessary for scaling the approach to larger circuits \cite{rodatz2025fault,poor2026ultra}. In this work, we address this limitation by developing an automated search procedure based on fault-equivalent rewrites.
Using this procedure, we discover a fault-tolerant circuit that outperforms the previously reported, manually designed construction.

Our main contributions are as follows:
\begin{itemize}
\item We develop an automated search procedure over fault-equivalent rewrites, including an enumeration of the circuits reachable under the selected rewrite rules and search depth bounds for each gadget considered (Section~\ref{sec:search}).

\item We apply the procedure to two case studies: Shor-style syndrome extraction for the $[[7,1,3]]$ Steane code (Section~\ref{sec:shor}) and a Steane-based dynamic syndrome-extraction protocol for the same code (Section~\ref{sec:flag}).
We also confirm that the manually derived circuit of Po{\'o}r \textit{et al.} lies within the search space induced by these rewrites  (Appendix~\ref{app:manual}) \cite{poor2026ultra}.

\item We perform circuit-level noise simulations showing that the selected circuits reduce the logical error rate relative to the corresponding baseline constructions for syndrome extraction with the $[[7,1,3]]$ Steane code (Section~\ref{sec:shorperf} and \ref{sec:steaneperf}).
\end{itemize}

The remainder of this paper is organized as follows. Section~\ref{sec:prelim} summarizes the notion of fault equivalence used in this work. Section~\ref{sec:rules} presents the rewrite rules and their constraints. Section~\ref{sec:search} describes the search algorithm. Sections~\ref{sec:shor} and~\ref{sec:flag} present the two case studies and the results of the corresponding circuit-level noise simulations.
\section{Preliminaries}
\label{sec:prelim}

This section introduces the fault-tolerance framework used throughout this work. We first state the fault-tolerance condition considered for syndrome-extraction gadgets. We then review fault equivalence and the edge-flip noise model for ZX diagrams, and describe how a circuit under circuit-level noise can be represented and transformed within this model. Finally, we introduce the two classes of syndrome-extraction gadgets studied in Sections~\ref{sec:shor} and~\ref{sec:flag}.

\subsection{Fault-tolerance condition}
\label{sec:ftdef}

Let $\mathcal{Q}$ be an $[[n,k,d]]$ stabilizer code with stabilizer group $\mathcal{S}$, and let

$$
    t = \left\lfloor \frac{d-1}{2} \right\rfloor
$$

be the number of errors that $\mathcal{Q}$ can correct. Since Pauli errors that differ by a stabilizer act identically on the
codespace~\cite{calderbank1998quantum,gottesman1997stabilizer}, we
define the stabilizer-coset weight of a Pauli error $E$ as
\begin{equation}
{wt}_{\mathcal{S}}(E)
=
\min_{s\in\mathcal{S}} {wt}(Es).
\label{eq:stabilizer-weight}
\end{equation}
Errors that differ by a stabilizer have the same action on the codespace. Consequently, it is the minimum weight within the stabilizer coset, rather than the physical weight of a particular error, that determines whether an error is correctable.

Consider a syndrome-extraction gadget $G$ acting on an initially error-free encoded state. Let $F$ denote a pattern of circuit faults, ${wt}(F)$ the number of fault events in that pattern, and $E_G(F)$ the resulting Pauli error on the data qubits. We say that $F$ is \emph{detectable} if it produces a nontrivial outcome in one or more of the gadget's designated verification measurements (flags) \cite{rodatz2025fault}. Otherwise, $F$ is \emph{undetectable}.

We adopt the standard condition for $t$-fault-tolerant syndrome extraction~\cite{chamberland2018flag}, denoted by $\mathrm{FT}_t$: 
\begin{equation}
\begin{aligned}
& (wt(F)\leq t)
\ \land
(F\text{ is undetectable})\\
& \quad\Longrightarrow\quad
{wt}_{\mathcal{S}}\left(E_G(F)\right)\leq wt(F).
\label{eq:ft-condition}
\end{aligned}
\end{equation}
Equivalently, no combination of at most $t$ undetectable faults may produce a data error whose weight modulo $\mathcal{S}$ exceeds the number of faults.

\subsection{Fault equivalence and edge-flip noise}
\label{sec:fault-equivalence}

Fault equivalence provides a relational characterization of circuits under noise \cite{rodatz2025fault}. Informally, two implementations are fault-equivalent if every undetectable fault in either implementation has a fault of no greater weight in the other implementation with the same effect.

More precisely, let circuits $C_1$ and $C_2$ implement the same ideal linear map under noise models $\mathcal{F}_1$ and $\mathcal{F}_2$, respectively. The circuits are $w$-fault-equivalent if, for every undetectable fault pattern of weight strictly less than $w$ in either circuit, there exists a fault pattern of no greater weight in the other circuit that produces the same faulty map. This correspondence is required in both directions. The circuits are fault-equivalent if they are $w$-fault-equivalent for every $w$. Because the weight bound in this definition is strict, preserving all fault patterns of weight at most $t$ corresponds to $(t+1)$-fault equivalence.

The ZX-calculus represents a quantum computation as a graph-like diagram and provides local rewrite rules for transforming that diagram without changing its ideal linear map \cite{coecke2008interacting,duncan2020graph}. Standard ZX rewrites, however, do not necessarily preserve the behavior of a computation under faults. To express faults without choosing a time direction for the diagram, Rodatz \textit{et al.} introduce the \emph{edge-flip noise model} \cite{rodatz2025fault}. Under this model, an atomic fault is a non-identity Pauli operator acting on a single edge of a ZX diagram. The weight of a fault is the minimum number of such edge flips required to represent it.
With these noise models in place, we can now define fault equivalence between circuits and ZX diagrams.
A ZX diagram $D$ under edge-flip noise is a \emph{fault-equivalent representation} of a circuit $C$ under a circuit-level noise model if $C$ and $D$ are fault-equivalent. General circuit faults, including correlated faults produced by multi-qubit gates, can be represented using fault gadgets in the ZX diagram \cite{rusch2025completeness}. In the Calderbank–Shor–Steane (CSS) codes considered in this paper, the $X$- and $Z$-error components are treated separately, allowing the circuit-level CSS noise model to be represented using the corresponding edge flips \cite{rodatz2025fault}.

\subsection{Fault-equivalent circuit transformation}
\label{sec:ft-transform}

To avoid the large branching factor of the ZX rewrite system, our optimization procedure represents search states as circuits rather than ZX diagrams. Each search transition performs the following steps
\begin{equation}
\begin{gathered}
C_i
\xrightarrow{\text{ZX translation}}
D_i
\xrightarrow{r_i}
\widetilde{D}_i
\xrightarrow{\text{circuit extraction}}
C_{i+1},
\\ (r_i\in\mathcal{R}_{\mathrm{FE}}),
\end{gathered}
\label{eq:transformation-pipeline}
\end{equation}
where $C_i$ is the circuit at the current search state, $D_i$ is its fault-equivalent ZX representation, $r_i$ is a single fault-equivalent rewrite instance selected from the rewrite set $\mathcal{R}_{\mathrm{FE}}$, $\widetilde{D}_i$ is the resulting ZX diagram, and $C_{i+1}$ is the circuit extracted from $\widetilde{D}_i$.

For each search state, the circuit-level noise model of $C_i$ is first mapped to edge-flip noise on $D_i$. The search enumerates the applicable fault-equivalent rewrite instances on $D_i$ and applies each instance separately. Rather than applying further ZX rewrites to the resulting diagram, the procedure immediately attempts to extract a circuit from $\widetilde{D}_i$. Diagrams that do not admit a valid circuit extraction are discarded, while each successfully extracted circuit $C_{i+1}$ becomes a successor state in the search. Repeating this procedure produces a sequence
\begin{equation}
C_0 \longrightarrow C_1 \longrightarrow \cdots
\longrightarrow C_m
\end{equation}
without directly exploring arbitrary sequences of rewrites in the substantially larger space of ZX diagrams.

The translation $C_i\to D_i$, the rewrite $D_i\to\widetilde{D}_i$, and the extraction $\widetilde{D}_i\to C_{i+1}$ each preserve fault equivalence. Therefore, $C_i$ and $C_{i+1}$ are fault-equivalent. By transitivity, every circuit reached by the search is fault-equivalent to the initial circuit $C_0$ and consequently retains its fault-tolerance properties. The rewrite rules and the conditions required for valid circuit extraction are presented in Section~\ref{sec:rules}.

\subsection{Gadgets under study}
\label{sec:gadgets}

\paragraph{Shor-style syndrome extraction.}
Shor-style syndrome extraction measures each stabilizer generator using a cat-state ancilla \cite{shor1996fault}. The ancilla is verified before interacting with the data so that faults introduced during its preparation do not propagate to an uncorrectable data error. Each qubit of the cat state interacts with at most one data qubit, thereby limiting the propagation of a single ancilla fault. Our first case study applies the automated rewrite search to the preparation, verification, and data-interaction circuits that constitute a complete round of Shor-style syndrome extraction for the $[[7,1,3]]$ Steane code. A raised verification flag is handled by decoding rather than by rejection. In the baseline circuit, the verification is complete before the cat state touches the data, so a flag there can mean \emph{discard the ancillas and re-prepare}, at a cost in time rather than in accuracy. A rewrite may move a data interaction ahead of the verification, after which that option is no longer available: by the time the flag is measured, the fault it reports may have already propagated to the data. Repeating the measurement does not help either, since a repeated round reports the resulting error faithfully and a weight-one decoder then corrects the wrong qubit. We therefore use one protocol for every circuit in this case study: when a flag fires, that round's syndrome is discarded in full and the following round is decoded through a lookup table conditioned on which gadget flagged, so that no run is rejected. Section~\ref{sec:shorp} states the protocol and gives the tables.

\paragraph{Steane-based dynamic syndrome extraction.}
Our second case study uses the syndrome-extraction circuit underlying the dynamic protocol of Po{\'o}r \textit{et al.} \cite{poor2026ultra}. The input circuit combines Goto's fault-tolerant preparation of the logical state $\ket{0}_L$ \cite{goto2016minimizing} with the transversal \textsc{cnot} coupling of Steane-style syndrome extraction \cite{steane1997active}. Goto's preparation circuit contains a single check qubit that verifies the encoded ancilla state. Within the dynamic protocol, the measurement of this check qubit serves as a flag: a nontrivial outcome indicates that a fault may have produced a correlated error in the encoded ancilla and triggers the recovery branch of the protocol.
\section{Rewrite rules}
\label{sec:rules}

This section introduces the fault-equivalent rewrite rules used by our search. Although the validity of these rules is established using ZX-calculus, we express them as coarse-grained circuit transformations. A single circuit-level rule can represent a sequence of elementary ZX rewrites, thereby avoiding the large branching factor that would result from searching directly over all intermediate ZX diagrams.

The symbol $\circumeq$ denotes fault equivalence under the noise models specified in Section~\ref{sec:prelim}. Thus, each rule preserves not only the ideal linear map but also the minimum weight and effect of undetectable faults. Each circuit-level rule is derived by translating the circuit on the left-hand side of the corresponding equation into a ZX diagram and then transforming the resulting diagram using fault-equivalent ZX rewrites. Detailed derivations are provided in Appendix~\ref{app:der_rewrite}.

\subsection{Restricted commutation}

The restricted-commutation rules exchange the order of two \textsc{cnot} gates that share an initialized qubit. In the first rule, the shared qubit is a control initialized in $\ket{+}$; in the second, it is a target initialized in $\ket{0}$. The initialization and the requirement that the other two \textsc{cnot} endpoints are distinct are part of the rules' applicability conditions.

\begin{equation}
    \includegraphics[]{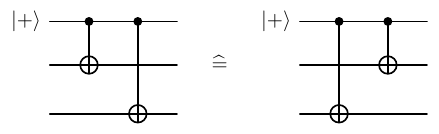}
\end{equation}
\begin{equation}
    \includegraphics[]{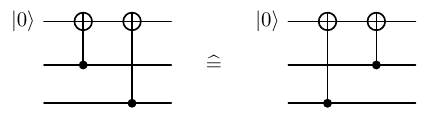}
\end{equation}
These transformations allow the search to change the temporal ordering of interactions without changing their fault-equivalence class. They cannot be applied when the shared qubit has an arbitrary input state.

\subsection{Basis swap}

The basis-swap rule uses the symmetry of Bell-state preparation. A \textsc{cnot} with a control initialized in $\ket{+}$ and a target initialized in $\ket{0}$ prepares $\ket{\Phi^+}$. Because this state is invariant under exchanging its two qubits, the roles and vertical positions of the two initialized wires may be exchanged:

\begin{equation}
    \includegraphics[]{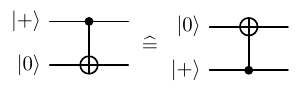}
\end{equation}
  
The initialization states are an essential constraint. In particular, this rule does not represent the insertion of a physical \textsc{swap} gate and does not apply to arbitrary input states.

\subsection{Target swap}

The target-swap rule changes the endpoint of the second \textsc{cnot} in the following three-qubit circuit:

\begin{equation}
    \includegraphics[]{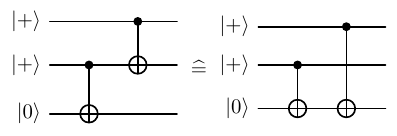}
\end{equation}

This rule preserves fault equivalence only for the indicated initialization pattern. It allows the search to redirect an interaction from one qubit to another while retaining the same encoded state and fault behavior.

\subsection{Bell reduction}

We define the Bell reduction of $(q_2,q_1)$ as the following transformation, which removes one qubit of a Bell pair:

\begin{equation}
    \includegraphics[]{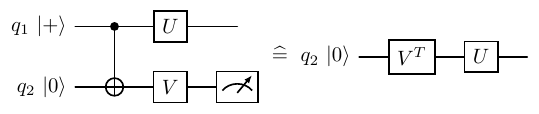}
\end{equation}

\paragraph{Constraints.}
\label{sec:validity}

The Bell-reduction rule has two constraints. First, after the Bell pair is prepared, its two qubits must evolve independently until the measurement; that is, their joint evolution must factor as $U\otimes V$. In particular, no subsequent gate may act jointly on the two qubits. Second, the outcome of the eliminated measurement must not be used independently as a flag or to control a subsequent operation. It may contribute only to a parity computed from several measurement outcomes.
Let $m$ denote the eliminated outcome and let $q$ be the parity of the remaining outcomes. If the protocol retains only
\begin{equation}
p=m\oplus q,
\end{equation}
then the records $(m,q)$ and $(0,m\oplus q)$ are operationally equivalent. We may therefore fix $m=0$ and absorb its value into the remaining parity bit \cite{poor2026ultra}. This is a relabeling of the classical measurement record, rather than physical postselection: runs in which $m=1$ are not discarded.

Under these conditions, the measurement can be represented by the outcome-$0$ effect $\bra{0}$. The transpose identity
\begin{equation}
(U\otimes V)\ket{\Phi^+}
=
\left(UV^{\mathsf T}\otimes I\right)\ket{\Phi^+}
\label{eq:bell-transpose}
\end{equation}
then gives
\begin{equation}
\left(I\otimes\bra{0}\right)
(U\otimes V)\ket{\Phi^+}
=
\frac{1}{\sqrt{2}}UV^{\mathsf T}\ket{0}.
\label{eq:bell-reduction}
\end{equation}
This is precisely the right-hand side of the Bell-reduction rule, up to the nonzero scalar $1/\sqrt{2}$. If either constraint is violated, the reduction does not generally apply.

\section{Search algorithm}
\label{sec:search}

The restricted-commutation, basis-swap, and target-swap rules do not directly reduce the resource cost of a circuit. Their purpose is to expose a subcircuit to which Bell reduction can be applied. Accordingly, our search does not treat applications of these enabling rules as independent search transitions. Instead, it combines a sequence of enabling rewrites with a subsequent Bell reduction into a single \emph{composite reduction}. 

Let
\begin{equation}
\mathcal{R}_{E}
=
\{
\textsc{Commute},
\textsc{BasisSwap},
\textsc{TargetSwap}
\}
\end{equation}
denote the set of enabling rules, and let $\mathcal{R}_{B}$ denote Bell reduction. A composite reduction has the form
\begin{equation}
C_i
\xrightarrow{\mathcal{R}_{E}^{*}}
\widetilde{C}_i
\xrightarrow{\mathcal{R}_{B}}
C_{i+1},
\label{eq}
\end{equation}
where the enabling sequence transforms $C_i$ into a circuit $\widetilde{C}_i$ containing a valid Bell-reduction pattern.
$\mathcal{R}_E^*$ denotes the set of all finite compositions of zero or more rules $r_i \in \mathcal{R}_E$, with the composition of zero rules defined as the identity.

The search treats the entire transformation
$C_i\to C_{i+1}$ as one atomic operation and does not add the intermediate circuits generated by $\mathcal{R}_{\mathrm{E}}^{*}$ to the search frontier.

For each circuit $C_i$, the algorithm identifies potential Bell-pair structures satisfying the constraints of Section~\ref{sec:rules}. For every such structure, it constructs a corresponding sequence of enabling rewrites, applies the Bell reduction, and adds the resulting circuit $C_{i+1}$ as a child of $C_i$. The procedure is then repeated recursively from each child.

\algrenewcommand{\algorithmicrequire}{\textbf{Input:}}
\algrenewcommand{\algorithmicensure}{\textbf{Output:}}

\begin{algorithm}[H]
\caption{Breadth-first composite-reduction search}
\label{alg}
\begin{algorithmic}[1]
\Require Circuit $C_0$
\Ensure Set of terminal circuits reachable from $C_0$
\State $\mathcal{Q}\gets [C_0]$ \Comment{FIFO queue}
\State $\mathcal{T}\gets\varnothing$
\While{$\mathcal{Q}$ is not empty}
    \State $C\gets\Call{Dequeue}{\mathcal{Q}}$
    \State $\mathit{terminal}\gets\mathrm{true}$
    \State $\mathcal{M}\gets\Call{FindCompositeReductions}{C}$
    \ForAll{$M\in\mathcal{M}$}
        \State $C'\gets\Call{ApplyCompositeReduction}{C,M}$
        \If{$C'$ is valid}
            \State $\mathit{terminal}\gets\mathrm{false}$
            \State $\Call{Enqueue}{\mathcal{Q},C'}$
        \EndIf
    \EndFor
    \If{$\mathit{terminal}$}
        \State $\mathcal{T}\gets\mathcal{T}\cup\{C\}$
    \EndIf
\EndWhile
\State \Return $\mathcal{T}$
\end{algorithmic}
\end{algorithm}
The implementation does not deduplicate circuits reached through different rewrite sequences. Consequently, the same circuit may occur at multiple nodes of the search tree and is processed independently along each path. This redundancy affects the running time but not the soundness of the search.

Every search transition contains exactly one Bell reduction, which removes an ancilla qubit and a \textsc{cnot} gate. The resource cost therefore decreases strictly along every root-to-leaf path. Assuming that the composite-reduction matcher returns finitely many matches for each circuit, the depth of the search tree is finite and bounded by the number of Bell reductions that can be applied to the input circuit.

Finally, every constituent transformation is fault-equivalent. By composition and transitivity, each composite reduction preserves fault equivalence, and every circuit returned by the search is fault-equivalent to the input circuit $C_0$.
\section{Evaluation}
\label{sec:evaluation}

\subsection{Simulation method}
\label{sec:simulations}

\paragraph{Noise models}

We use Stim~\cite{gidney2021stim} for circuit-level noise simulation and a lookup-table decoder for the $[[7,1,3]]$ code. We denote the two-qubit gate error rate by $p$ and set the single-qubit gate error rate to $p_1=3p/100$ and the state-preparation and measurement error rate to $p_{\mathrm{SPAM}}=p$, following Ref.~\cite{poor2026ultra}. These relative error rates are motivated by the Quantinuum H2 processor~\cite{decross2025computational}.
For idle noise, we use the depolarizing model with the error probability of $p/10$ by default. However, for comparison with previous work, we consider three idle-noise models for the Steane-based gadget. The dephasing model introduces a $Z$ error with probability $p/10$  \cite{poor2026ultra}. The two depolarizing models assign probabilities of $p/30$ and $p/10$, respectively, to each of the Pauli errors $X$, $Y$, and $Z$. Thus, the first depolarizing model has the same total idle-error probability as the dephasing model, whereas the second has the same $Z$-error probability.
We perform two parameter scans. First, we vary the number of QEC cycles from $N=1$ to $10$ at
fixed physical error rate $p=10^{-3}$. Second, we vary $p$ from $10^{-4}$ to $10^{-2}$ with
$N=1$.

\subsubsection{Logical-error estimation}

We estimate logical error rates by ideally preparing either $\ket{0}_L$ or $\ket{+}_L$, applying
the specified number of noisy QEC cycles, and performing ideal final readout followed by
lookup-table decoding. A logical error occurs when the decoded logical observable differs from
that of the initially prepared state.

\subsection{Shor-style syndrome extraction}
\label{sec:shor}
\label{sec:shorft}

\subsubsection{Baseline protocol and search space}

A gadget measures one stabilizer generator: it prepares a $w$-qubit cat state on fresh ancillas,
applies a controlled-$P$ from ancilla $j$ to the $j$-th data qubit of the support, and measures the
ancillas in the $X$ basis. The parity of the ancilla measurements determines the eigenvalue of the generator. The
ancilla is the control for both generator types; a $Z$-type gadget is the $X$-type one conjugated
by Hadamards on the data, that is, a controlled-$Z$.

\begin{figure}[t]
\centering
  \includegraphics[width=0.48\textwidth]{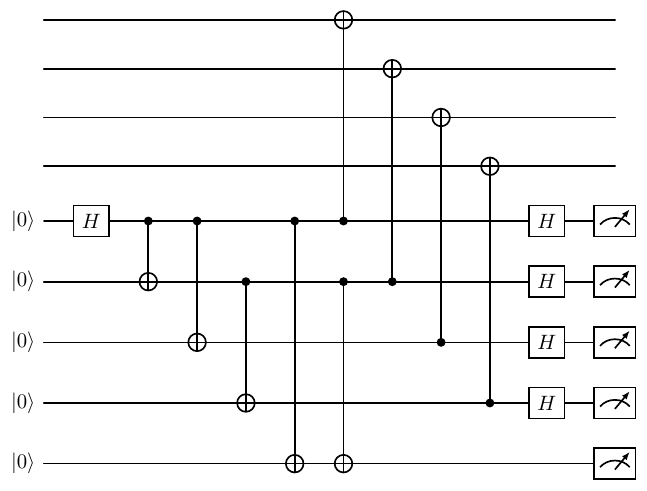}
  \smallskip
\textbf{(a)} Baseline circuit\par
\medskip
  \includegraphics[width=0.48\textwidth]{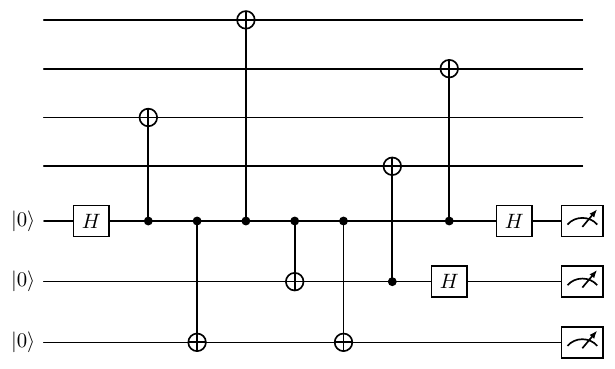}
  \smallskip
\textbf{(b)} Optimized circuit\par
\medskip
\caption{Shor-style gadgets for measuring a weight-four $X$-type
stabilizer generator: (a) the baseline circuit and (b) the optimized
circuit. In each panel, the upper four wires are data qubits, the
intermediate wires are ancillas, and the bottom wire is a flag qubit.
The derivation of the reduced circuit is given in
Appendix~\ref{app:shor}.}
\label{fig:shor-circuits}
\end{figure}

The baseline prepares each cat state using a tree of \textsc{cnot}s and verifies it using a single parity check. This circuit has been identified as the best among the zero-, one-, and two-check variants~\cite{weinstein2012shor,tomita2013comparison}. Per generator, this configuration requires five ancillas, nine \textsc{cnot}s, and a \textsc{cnot} depth of five. A complete round over all six generators uses 30 ancilla preparations, 54 \textsc{cnot}s, and six verification flags. Fault enumeration confirms that the baseline gadget is $FT_1$.

Subject to several constraints—most notably that a verification flag is never merged into the preparation circuit—we exhaustively enumerate the circuits reachable via the rewrites in Section~\ref{sec:rules}. This yields seven circuits per gadget, with identical structures for the $Z$- and $X$-measurement cases.

\subsubsection{Search results}

\begin{table}[ht]
\centering
\caption{Resource cost of Shor gadget for the $[[7,1,3]]$ code before and after the reduction. Depth is \textsc{cnot}
depth. Both generator types are reduced identically using the Bell reductions of $(7,9)$ and $(7,8)$.}
\label{tab:shor-gadget}
\begin{tabular}{lrrr}
\toprule
circuit & ancillas & \textsc{cnot}s & depth \\
\midrule
baseline & 5 & 9 & 5 \\
reduced  & 3 & 7 & 6 \\
\bottomrule
\end{tabular}
\end{table}

Of the seven circuits the search returns for each gadget, we choose the one with the fewest ancillas and, among those, the least CNOT depth. Monte Carlo simulation of the candidates under the protocol of Section~\ref{sec:shorp} confirms that this circuit also has the lowest logical error rate. As Table~\ref{tab:shor-gadget} shows, the chosen circuit reduces each gadget from five to three ancilla preparations and from nine to seven CNOTs, at the cost of one additional layer of CNOT depth. Consequently, an optimized round requires 18 ancilla preparations and 42 \textsc{cnot}s, compared with 30 and 54, respectively, for the baseline, while retaining all six verification flags.

\subsubsection{Repeated-syndrome protocol}
\label{sec:shorp}
A cycle is three rounds of syndrome measurements followed by one decoding step, using Shor's rule: take the first syndrome
that repeats in two consecutive rounds, otherwise the last one measured. 

A correction is decoded once per cycle and accumulated into a Pauli frame carried into the next
cycle, so the decoder always acts on the difference between the reported syndrome and what the
frame already accounts for.

Raised verification flags are handled by repeating a round using the same gadgets. The decoding table for each gadget is constructed by enumerating all single faults, retaining those that raise the gadget’s flag, and mapping the syndrome of each residual error to its corresponding correction. Residual errors are projected onto the basis to which the memory is sensitive and considered modulo the stabilizer group. Each table is single-valued, so every syndrome has a well-defined correction. This projection is only a simulation convenience and does not exploit knowledge of the prepared state. Even without the projection, when the full syndrome is retained and residual errors are compared modulo the stabilizer group, the tables remain unambiguous. Thus, a decoder that simultaneously corrects both error types can use them unchanged.

Table~\ref{tab:flag-table} gives the lookup-table decoder for the $\ket{0}_L$ memory. The baseline $Z$-type gadgets require no table: in a baseline $Z$-type gadget, an internal $X$ error raises the flag but can induce only a $Z$ error on a data qubit, to which the $\ket{0}_L$ memory is insensitive. The reduced $Z$-type gadgets do not share this symmetry: four of their 33 single faults escape the gadget and leave a weight-one $X$ residual, accounting for the sole nontrivial entry in their decoding tables.

The same decoding rule applies to every table entry. When a flag is raised in a round, that round’s syndrome is discarded entirely. The syndrome from the next round is then looked up in the table corresponding to the gadget that raised the flag, and the resulting correction is applied to the Pauli frame. If a flag is raised in the very last round of the experiment, the perfect readout is used instead of a subsequent syndrome.

\begin{table*}[!p]
\centering
    \caption{Flag-conditioned decoding tables for the $\ket{0}_L$ memory. For each gadget, the support of the stabilizer generator it measures is specified as a set of qubit indices. Syndromes are written $s_3 s_2 s_1$, where $s_1$, $s_2$ and $s_3$ are the outcomes of $Z\{0,1,2,3\}$, $Z\{1,2,4,5\}$ and $Z\{2,3,5,6\}$, so the rightmost bit belongs to $Z\{0,1,2,3\}$.The corresponding tables for the $\ket{+}_L$ memory are obtained by exchanging $X$ and $Z$.}
\label{tab:flag-table}
\begin{tabular}{llcl}
\toprule
circuit & gadget & faults raising the flag & Z syndrome $\to$ correction \\
\midrule
baseline  & $X\{0,1,2,3\}$ & 42 & $001\!\to\!X_0$, $011\!\to\!X_1$, $110\!\to\!X_1X_3$ \\
          & $X\{1,2,4,5\}$ & 42 & $001\!\to\!X_2X_5$, $011\!\to\!X_1$, $111\!\to\!X_2$ \\
          & $X\{2,3,5,6\}$ & 42 & $001\!\to\!X_3X_6$, $101\!\to\!X_3$, $111\!\to\!X_2$ \\
          & $Z$-type       & 42 & None \\
\midrule
reduced   & $X\{0,1,2,3\}$ & 33 & $011\!\to\!X_1$, $110\!\to\!X_1X_3$, $111\!\to\!X_0X_1X_3$ \\
          & $X\{1,2,4,5\}$ & 33 & $001\!\to\!X_2X_5$, $010\!\to\!X_1X_2X_5$, $111\!\to\!X_2$ \\
          & $X\{2,3,5,6\}$ & 33 & $001\!\to\!X_3X_6$, $101\!\to\!X_3$, $110\!\to\!X_2X_3X_6$ \\
          & $Z\{0,1,2,3\}$ & 33 & $001\!\to\!X_0$ \\
          & $Z\{1,2,4,5\}$ & 33 & $011\!\to\!X_1$ \\
          & $Z\{2,3,5,6\}$ & 33 & $111\!\to\!X_2$ \\
\bottomrule
\end{tabular}
\end{table*}

\subsubsection{Logical-error performance}
\label{sec:shorperf}
At $N=10$ cycles, replacing the baseline gadget with the reduced gadget lowers the logical error rate from $4.23 \times 10^{-2}$ to $3.45 \times 10^{-2}$ in the
$Z$ basis and from $4.03 \times 10^{-2}$ to $3.37 \times 10^{-2}$ in the $X$ basis. The agreement between the ratios for the two memories is both a result and a consistency check: a self-dual code with symmetric gadgets should exhibit no basis preference. Fits over $p \leq 2 \times 10^{-3}$ yield log--log slopes ranging from $1.94$ to $1.97$ across both circuits and both memories. This near-quadratic scaling confirms that the flag decoder has no first-order failure mode: if a decoding table miscorrected even one single fault, the corresponding slope would be near one.

Flags fire less often in the reduced round --- $0.062$ against $0.072$ per cycle at $p = 10^{-3}$,
and $0.47$ against $0.53$ at ten cycles --- because there are fewer gates upstream of each flag.
Since flags are decoded rather than post-selected this is not a yield.

Both circuits are scored under the same protocol: a flag selects a decoding table, no shot is discarded, and each circuit
recovers with its own gadgets --- the round after the flagged one is another copy of the same
circuit. This is also what the reference circuits of Section~\ref{sec:flag} do,
whose decoder keeps every shot, discard the flagged round's syndrome and reads the following
round through a flag-conditioned lookup table, so the two families in this paper are scored the same way.

\begin{figure*}[!p]
\centering
\includegraphics[width=\textwidth]{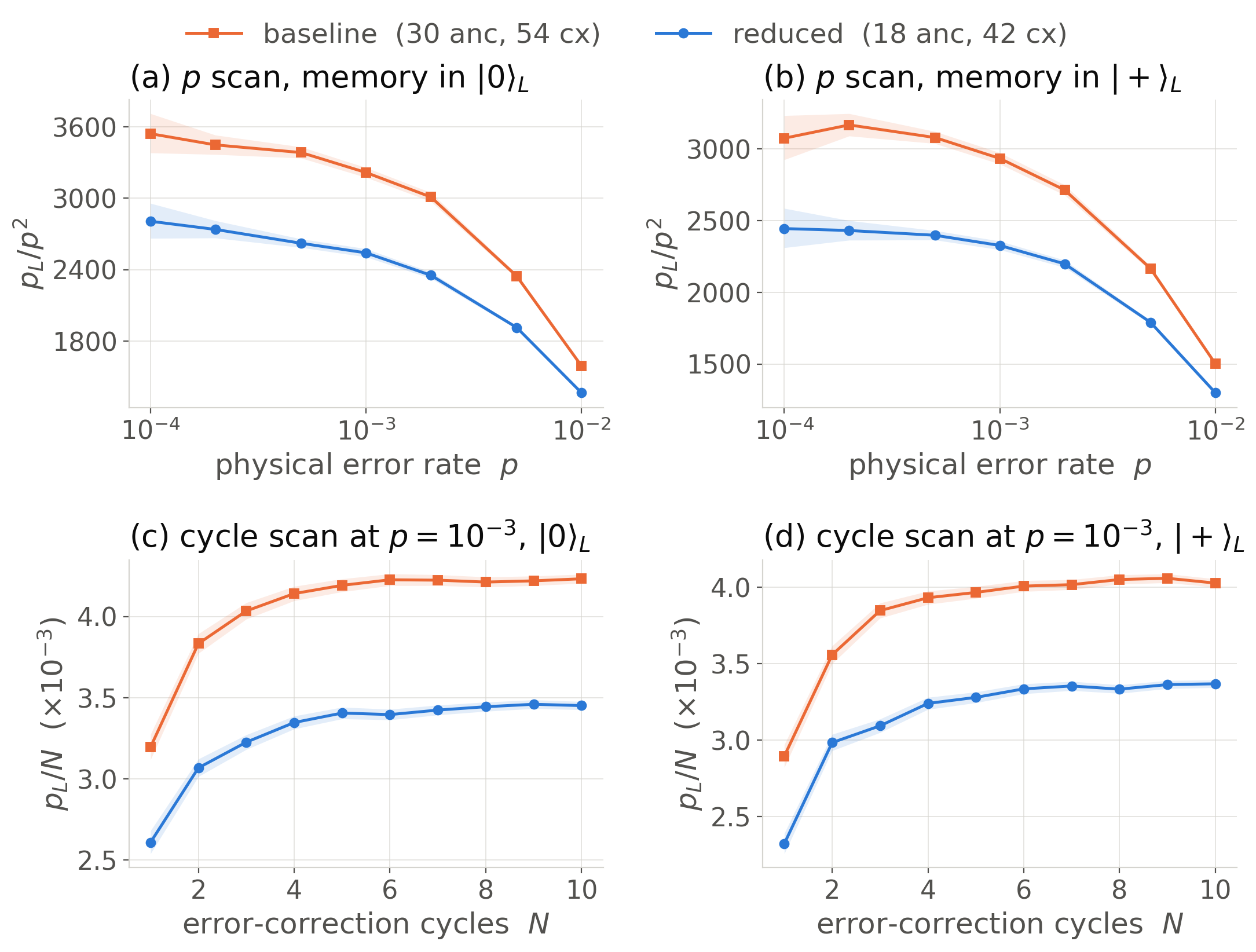}
\caption{Logical error rates for the baseline and optimized Shor-style syndrome-extraction circuits. Each QEC cycle comprises three rounds of syndrome measurement followed by a decoding step. Panels (a) and (b) vary the physical two-qubit gate error rate $p$ for a single QEC cycle ($N=1$) and plot $p_L/p^2$, thereby removing the common quadratic dependence on $p$ and isolating the leading-order coefficient. Panels (c) and (d) vary the number of cycles $N$ at $p=10^{-3}$ and plot the per-cycle logical error rate $p_L/N$, which remains flat as a function of $N$ for $N \geq 5$. Results are shown for the $\ket{0}_L$ and $\ket{+}_L$ memories. A raised verification flag invokes an alternative decoding table rather than causing the run to be rejected; consequently, no shots are discarded, and the reported rates are unconditional. Bands show $95\%$ Wilson confidence intervals.}
\label{fig:shor-schemes}
\end{figure*}

Figure~\ref{fig:shor-schemes} compares the two circuits. Panels (a) and (b) plot the logical error rate normalized by $p^2$. Because both circuits exhibit quadratic scaling, this normalization isolates the leading-order coefficient, which distinguishes their performance. Panels (c) and (d) plot the error rate per cycle, which remains flat for both circuits when $N \geq 5$.

\subsection{Steane-based dynamic syndrome extraction}
\label{sec:flag}

\subsubsection{Baseline protocol and search space}

The baseline is the syndrome-measurement circuit for the $[[7,1,3]]$ code constructed from
Goto's fault-tolerant $\ket{0}_L$ preparation \cite{goto2016minimizing}. In the resulting dynamic protocol,
Goto's verification qubit serves as the flag.

The search is applied to the primary circuit only. When the flag fires, the protocol branches to
the recovery circuit of Po{\'o}r \textit{et al.}~\cite{poor2026ultra} --- eleven \textsc{cnot}s on three
ancillas, not itself fault-tolerant --- and we leave that branch untouched, so the searched circuit
and the circuit it is compared against differ only in the primary circuit. The flag
conditions the decoding in the same way for both: the flagged round's syndrome is discarded, the
recovery circuit is run, and its correction is read through the flag-conditioned lookup table of the dynamic
protocol.

Using composite reductions formed from the enabling rules and Bell reduction, the search generated 391 candidate circuits. We evaluated these candidates using Monte Carlo simulation and selected the circuit with the lowest estimated logical error rate.

\subsubsection{Search results}

\begin{figure}[!tp]
\centering
  \includegraphics[width=0.48\textwidth]{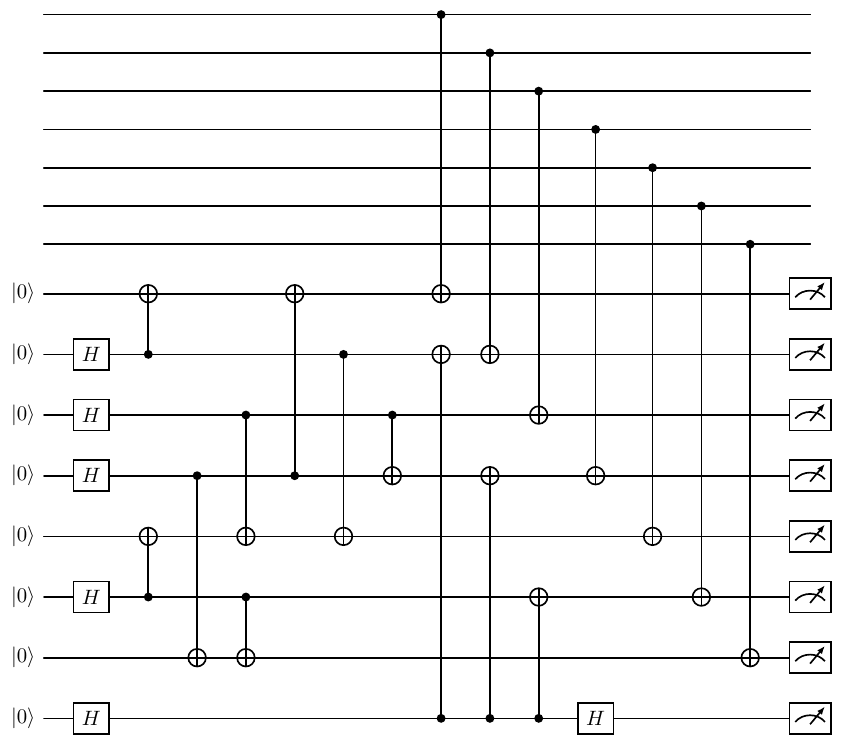}
  \smallskip
\textbf{(a)} Baseline circuit\par
\medskip
  \includegraphics[width=0.48\textwidth]{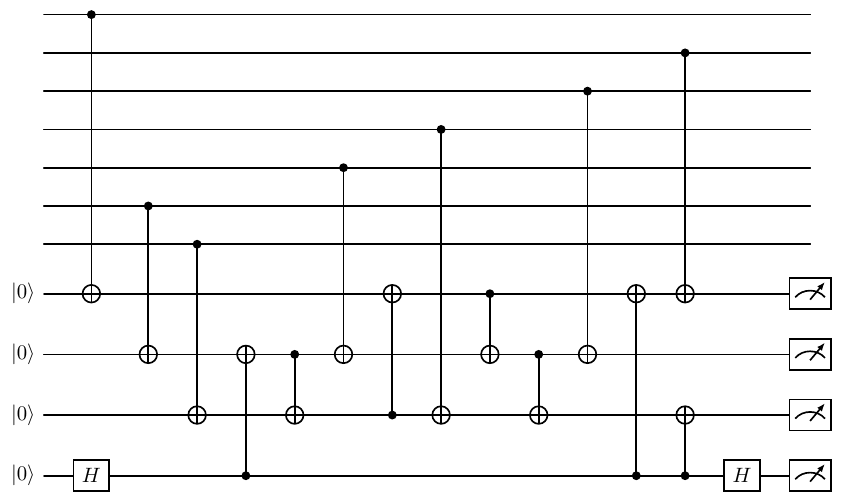}
  \smallskip
\textbf{(b)} Optimized circuit\par
\medskip
\caption{Steane error-correction circuits: (a) the baseline circuit based on Goto's construction~\cite{goto2016minimizing} and (b) the optimized circuit found by our search. Qubits $0,\ldots,6$ are data qubits. Four Bell reductions transform the baseline into the optimized circuit, reducing the number of ancillas from eight to four, \textsc{cnot} gates from 18 to 14, and Hadamard gates from six to two.}\label{fig:steane}
\end{figure}

To suppress branching, enabling rewrites are applied only when they immediately expose a Bell pair; they are never chained speculatively in anticipation of a later reduction. Consequently, each search step removes one ancilla, bounding the search-tree depth by the number of Bell reductions admitted by the input circuit. The trade-off is that the search excludes circuits reachable only through longer sequences of enabling rewrites.

The search generates 391 circuits from the Goto baseline. These circuits are ranked through a sequential race: each is evaluated in a ten-cycle memory experiment at progressively larger shot counts ($10^3$, $5 \times 10^3$, $2.5 \times 10^4$, and $10^5$). At each stage, circuits that fall behind are eliminated using a significance threshold of $\alpha=0.05$, and the survivors are finally evaluated with $10^6$ shots. The race is conducted on the $\ket{+}_L$ memory under the dephasing idle-noise model because it is the more challenging basis under this model. Of the 391 circuits, 27 survive.The winning circuit, obtained through the sequence of Bell reductions $(7,8)$, $(10,8)$, $(10,9)$, and $(10,8)$, is shown in Fig.~\ref{fig:steane}. Its derivation is given in Appendix~\ref{appendix:goto-derivation}.
\begin{table}[ht]
\centering
\caption{Resource comparison of the circuit found by our search with the baseline from which it was derived and a previously proposed circuit over one syndrome-extraction half-cycle. Total depth is computed using greedy scheduling, treating each Hadamard and \textsc{cnot} as a unit-duration gate. Resets are excluded because every circuit performs a single parallel reset of all ancillas before any other operation. The \textsc{cnot} depth is the number of parallel \textsc{cnot} rounds, with two \textsc{cnot}s sharing a layer if and only if they act on disjoint qubits. This quantity determines how long the data block remains idle.}
\label{tab:goto-resources}
\begin{tabular}{lrrrrr}
\toprule
circuit & ancillas & \textsc{cnot}s & $H$ & depth & \textsc{cnot} depth \\
\midrule
\makecell{Goto baseline\\(search input)} & 8 & 18 & 6 & 8 & 7 \\
\makecell{Dynamic Optimized\\Steane~\cite{poor2026ultra}} & 4 & 14 & 2 & 8 & 8 \\
\makecell{Searched\\(this work)} & 4 & 14 & 2 & 8 & \textbf{7} \\
\bottomrule
\end{tabular}
\end{table}

Relative to the Goto baseline, the reduction halves the ancilla count from eight to four and eliminates four \textsc{cnot}s and four Hadamards. The comparison with the Po{\'o}r circuit isolates the scheduling advantage: the reduced and Po{\'o}r circuits have identical resource counts—four ancillas, fourteen \textsc{cnot}s, and two Hadamards—and the same total depth of eight, but their \textsc{cnot} depths are seven and eight, respectively. Thus, the reduction provides no gate-count advantage in this comparison; its benefit lies entirely in scheduling.

Seven is a lower bound on the \textsc{cnot} depth of both circuits. In each circuit, one ancilla participates in seven \textsc{cnot} gates. Because gates acting on the same qubit cannot occupy the same layer, these gates require at least seven layers. The reduced circuit attains this bound.

\subsubsection{Logical-error performance}
\label{sec:steaneperf}
\begin{table*}[!tp]
\centering
\caption{Logical failure probability normalized by the number of cycles, $p_L/N$, after $N=10$
syndrome-extraction cycles at $p_2=10^{-3}$. Values are in units of $10^{-4}$; the best result in
each column is shown in bold. The three idle-noise models differ only in the channel applied to idling
qubits: dephasing at $p/10$, the model of \cite{poor2026ultra}; depolarizing at the same rate,
whose dephasing component is three times smaller; and depolarizing at $3p/10$, which matches that
dephasing component while also costing $X$ and $Y$ errors. Under dephasing every circuit is
markedly cheaper in the $Z$ basis than the $X$ basis; under either depolarizing model that
asymmetry disappears. The $95\%$ Wilson intervals are within $\pm1.4\%$ of every value.}
\label{tab:steane-comparison}
\begin{tabular}{lccccccc}
\toprule
& & \multicolumn{2}{c}{dephasing $p/10$}
& \multicolumn{2}{c}{depolarizing $p/10$}
& \multicolumn{2}{c}{depolarizing $3p/10$} \\
\cmidrule(lr){3-4}
\cmidrule(lr){5-6}
\cmidrule(lr){7-8}
circuit & ancillas & $Z$ basis & $X$ basis & $Z$ basis & $X$ basis & $Z$ basis & $X$ basis \\
\midrule
Three-qubit~\cite{reichardt2021fault}
& 3 & 3.58 & 6.82 & 4.74 & 5.69 & \textbf{9.31} & 11.58 \\
Steane~\cite{steane1997active}
& 8 & 3.59 & 8.58 & 6.68 & 6.55 & 14.97 & 14.99 \\
Optimized Steane~\cite{rodatz2025fault}
& 5 & 2.98 & 7.37 & 5.65 & 5.65 & 13.31 & 13.33 \\
Dynamic optimized Steane~\cite{poor2026ultra}
& 4 & 2.51 & 6.34 & 4.83 & 4.80 & 11.56 & 11.49 \\
Searched circuit (this work)
& 4 & \textbf{2.37} & \textbf{5.54} & \textbf{4.37} & \textbf{4.28} & 9.78 & \textbf{9.74} \\
\bottomrule
\end{tabular}
\end{table*}

Table~\ref{tab:steane-comparison} compares the searched circuit with the reference circuits under
all three idle-noise models. Under dephasing idle-noise every circuit has a markedly lower logical
error rate for the $\ket{0}_L$ memory than for the $\ket{+}_L$ memory, because dephasing noise
preferentially affects the latter. That asymmetry disappears under either depolarizing model for
every circuit but the three-qubit one, which keeps a $Z$ preference of $17\%$ at $p/10$ and $24\%$
at $3p/10$. The searched circuit achieves the lowest logical error rate in five of the six columns;
in the remaining one, the $Z$ basis under depolarizing $3p/10$, it is $5\%$ above the three-qubit
circuit.

Shot counts differ by point, because several of these runs allocate shots adaptively to a target
number of logical events rather than to a fixed budget. Under dephasing idle noise the reference
circuits use $10^7$ shots per point and the searched circuit $3.2 \times 10^7$; under depolarizing
idle noise at $p/10$ the reference circuits use $3.4$--$4.5 \times 10^6$ and the searched circuit
$8 \times 10^6$; at $3p/10$ every circuit uses $2 \times 10^6$. The resulting $95\%$ Wilson
intervals are between $0.46\%$ and $1.43\%$ of the value, so no comparison in the table rests on
overlapping intervals except the one named above.

Because the published reference results use dephasing idle noise, they cannot be compared directly
with simulations employing depolarizing idle noise. We therefore re-simulated all reference
circuits under both depolarizing models. Detailed parameter scans, over both the physical error rate and the number of cycles and for all three idle-noise models, are given in Appendix~\ref{app:data}.

\section{Conclusion and outlook}
\label{sec:conclusion}

We introduced an automated search for reducing fault-tolerant circuits using composite
fault-equivalent reductions. The enabling rewrite rules expose Bell-pair reductions, each of which
removes an ancilla preparation and a \textsc{cnot}. Because every transformation is fault-equivalent, the
resulting circuits retain the fault-tolerance properties of the input circuit by construction.

Applied to Shor-style syndrome extraction for the $[[7,1,3]]$ code, the search reduces one
complete syndrome-measurement round from 30 to 18 ancilla preparations and from 54 to 42 \textsc{cnot}s,
corresponding to reductions of $40\%$ and $22\%$, respectively. Across the simulated parameter
ranges, the optimized circuit reduces the logical error rate by $13$--$23\%$, and by approximately
$21\%$ at $p = 10^{-3}$ in both memories. Verification flags also fire less often: after ten QEC
cycles, for example, the fraction of runs in which some flag is raised falls from $0.527$ to
$0.472$.

For Steane-based dynamic syndrome extraction, the search generated 391 candidate circuits. The
selected circuit uses four ancillas and 14 \textsc{cnot}s, matching the resource counts of the circuit of
Po{\'o}r \textit{et al.}~\cite{poor2026ultra}, while achieving a lower estimated logical error rate.
Under depolarizing idle noise at $3p/10$ the reduction is approximately $15\%$ for both logical
basis states, and approximately $10\%$ at $p/10$. What separates the two circuits is not their
counts but their schedule: the selected circuit has a \textsc{cnot} depth of seven against eight, and idle
noise is what prices that layer. These results indicate that fault-equivalent rewriting can improve
circuit performance even when it does not reduce the gate and ancilla counts.

A limitation of the present implementation is that candidate performance is evaluated using
Monte Carlo simulation. Such estimates are computationally expensive and statistically noisy,
making them unsuitable as an objective for pruning every intermediate search branch. Circuit
depth alone is also insufficient as a performance proxy: in the Shor-style case, the \textsc{cnot} depth
increases from five to six even though the logical error rate decreases. A more promising search
objective would combine the number and type of fault locations, idle exposure, verification
acceptance, and a low-order count of malignant fault combinations.

Further improvements could incorporate circuit canonicalization and deduplication, heuristic or
beam-guided exploration, and independent high-precision evaluation of the most promising
candidates. Other directions include tracking the classical outcome maps induced by rewrites,
incorporating hardware connectivity and scheduling constraints, and extending the search to
higher-distance codes and other fault-tolerant gadgets. The present search explores only circuits
reachable through the chosen rewrite rules and therefore does not establish global optimality;
expanding the rewrite set while controlling its branching factor remains an important direction
for future work.

\begin{acknowledgments}
This work was supported by the National Research Foundation of Korea (NRF) Grant No. RS-2024-00442855 and No. RS-2024-00413957, all funded by the Korean government (MSIT).

The data supporting the findings of this study are available on GitHub~\cite{github}.
\end{acknowledgments}

\bibliographystyle{iopart-num-long}
\bibliography{references}

@article{aliferis2005quantum,
  title={Quantum accuracy threshold for concatenated distance-3 codes},
  author={Aliferis, Panos and Gottesman, Daniel and Preskill, John},
  journal={arXiv preprint quant-ph/0504218},
  year={2005},
  url={https://arxiv.org/abs/quant-ph/0504218}
}

@misc{github,
  title        = {Implementation and data for {automated reduction of fault-tolerant circuits}},
  howpublished = {\url{https://github.com/ijshj10/ftzx}}
}

@article{rusch2025completeness,
  title={{Completeness for fault equivalence of Clifford ZX diagrams}},
  author={R{\"u}sch, Maximilian and Kissinger, Aleks and Rodatz, Benjamin},
  journal={arXiv preprint arXiv:2510.08477},
  year={2025},
  url={https://arxiv.org/abs/2510.08477}
}

@inproceedings{shor1996fault,
  title={Fault-tolerant quantum computation},
  author={Shor, Peter W},
  booktitle={Proceedings of 37th conference on foundations of computer science},
  pages={56--65},
  year={1996},
  organization={IEEE},
  url={https://doi.org/10.1109/SFCS.1996.548464}
}

@article{steane1997active,
  title={Active stabilization, quantum computation, and quantum state synthesis},
  author={Steane, Andrew M},
  journal={Physical Review Letters},
  volume={78},
  number={11},
  pages={2252},
  year={1997},
  publisher={APS},
  url={https://doi.org/10.1103/PhysRevLett.78.2252}
}

@article{rodatz2025fault,
  title={Fault tolerance by construction},
  author={Rodatz, Benjamin and Po{\'o}r, Boldizs{\'a}r and Kissinger, Aleks},
  journal={arXiv preprint arXiv:2506.17181},
  year={2025},
  url={https://arxiv.org/abs/2506.17181}
}

@article{poor2026ultra,
  title={Ultra low overhead syndrome extraction for the Steane code},
  author={Po{\'o}r, Boldizs{\'a}r and Rodatz, Benjamin and Kissinger, Aleks},
  journal={Quantum Science and Technology},
  volume={11},
  number={4},
  pages={045009},
  year={2026},
  publisher={IOP Publishing},
  url={https://doi.org/10.1088/2058-9565/ae8df6}
}

@article{chao2018quantum,
  title={Quantum error correction with only two extra qubits},
  author={Chao, Rui and Reichardt, Ben W},
  journal={Physical Review Letters},
  volume={121},
  number={5},
  pages={050502},
  year={2018},
  publisher={APS},
  url={https://doi.org/10.1103/PhysRevLett.121.050502}
}

@article{reichardt2021fault,
  title={Fault-tolerant quantum error correction for {Steane’s} seven-qubit color code with few or no extra qubits},
  author={Reichardt, Ben W},
  journal={Quantum Science and Technology},
  volume={6},
  number={1},
  pages={015007},
  year={2021},
  publisher={IOP Publishing},
  url={https://iopscience.iop.org/article/10.1088/2058-9565/abc6f4/meta}
}

@article{zen2025quantum,
  title={Quantum circuit discovery for fault-tolerant logical state preparation with reinforcement learning},
  author={Zen, Remmy and Olle, Jan and Colmenarez, Luis and Puviani, Matteo and M{\"u}ller, Markus and Marquardt, Florian},
  journal={Physical Review X},
  volume={15},
  number={4},
  pages={041012},
  year={2025},
  publisher={APS},
  url={https://doi.org/10.1103/gqpr-dgz7}
}

@article{peham2025automated,
  title={Automated synthesis of fault-tolerant state preparation circuits for quantum error-correction codes},
  author={Peham, Tom and Schmid, Ludwig and Berent, Lucas and M{\"u}ller, Markus and Wille, Robert},
  journal={PRX Quantum},
  volume={6},
  number={2},
  pages={020330},
  year={2025},
  publisher={APS},
  url={https://doi.org/10.1103/PRXQuantum.6.020330}
}

@article{coecke2008interacting,
  title={Interacting Quantum Observables, Automata, Languages and Programming},
  author={Coecke, Bob and Duncan, Ross},
  journal={Lecture Notes in Computer Science},
  volume={5126},
  pages={298--310},
  year={2008}
}

@inproceedings{chen2025verifying,
  title={Verifying fault-tolerance of quantum error correction codes},
  author={Chen, Kean and Liu, Yuhao and Fang, Wang and Paykin, Jennifer and Wu, Xin-Chuan and Schmitz, Albert and Zdancewic, Steve and Li, Gushu},
  booktitle={International Conference on Computer Aided Verification},
  pages={3--27},
  year={2025},
  organization={Springer},
  url={https://doi.org/10.1007/978-3-031-98685-7_1}
}

@article{chamberland2018flag,
  title={Flag fault-tolerant error correction with arbitrary distance codes},
  author={Chamberland, Christopher and Beverland, Michael E},
  journal={Quantum},
  volume={2},
  pages={53},
  year={2018},
  publisher={Verein zur F{\"o}rderung des Open Access Publizierens in den Quantenwissenschaften},
  url={https://doi.org/10.22331/q-2018-02-08-53}
}

@article{chao2020flag,
  title={Flag fault-tolerant error correction for any stabilizer code},
  author={Chao, Rui and Reichardt, Ben W},
  journal={PRX Quantum},
  volume={1},
  number={1},
  pages={010302},
  year={2020},
  publisher={APS},
  url={https://doi.org/10.1103/PRXQuantum.1.010302}
}

@article{shutty2022decoding,
  title={Decoding merged color-surface codes and finding fault-tolerant {Clifford} circuits using solvers for satisfiability modulo theories},
  author={Shutty, Noah and Chamberland, Christopher},
  journal={Physical Review Applied},
  volume={18},
  number={1},
  pages={014072},
  year={2022},
  publisher={APS},
  url={https://doi.org/10.1103/PhysRevApplied.18.014072}
}

@article{rodatz2026floquetifying,
  title={Floquetifying stabiliser codes with distance-preserving rewrites},
  author={Rodatz, Benjamin and Po{\'o}r, Boldizs{\'a}r and Kissinger, Aleks},
  journal={Quantum},
  volume={10},
  pages={2202},
  year={2026},
  publisher={Verein zur F{\"o}rderung des Open Access Publizierens in den Quantenwissenschaften},
  url={https://doi.org/10.22331/q-2026-09-03-2202}
}

@article{duncan2020graph,
  title={{Graph-theoretic simplification of quantum circuits with the ZX-calculus}},
  author={Duncan, Ross and Kissinger, Aleks and Perdrix, Simon and Van De Wetering, John},
  journal={Quantum},
  volume={4},
  pages={279},
  year={2020},
  publisher={Verein zur F{\"o}rderung des Open Access Publizierens in den Quantenwissenschaften},
  url={https://doi.org/10.22331/q-2020-06-04-279}
}

@article{Prabhu2023faulttolerant,
  doi = {10.22331/q-2023-10-24-1154},
  url = {https://doi.org/10.22331/q-2023-10-24-1154},
  title = {Fault-tolerant syndrome extraction and cat state preparation with fewer qubits},
  author = {Prabhu, Prithviraj and Reichardt, Ben W.},
  journal = {{Quantum}},
  issn = {2521-327X},
  publisher = {{Verein zur F{\"{o}}rderung des Open Access Publizierens in den Quantenwissenschaften}},
  volume = {7},
  pages = {1154},
  month = oct,
  year = {2023}
}

@article{goto2016minimizing,
  title={{Minimizing resource overheads for fault-tolerant preparation of encoded states of the Steane code}},
  author={Goto, Hayato},
  journal={Scientific Reports},
  volume={6},
  number={1},
  pages={19578},
  year={2016},
  publisher={Nature Publishing Group UK London},
  url={https://www.nature.com/articles/srep19578}
}

@article{decross2025computational,
  title={Computational power of random quantum circuits in arbitrary geometries},
  author={DeCross, Matthew and Haghshenas, Reza and Liu, Minzhao and Rinaldi, Enrico and Gray, Johnnie and Alexeev, Yuri and Baldwin, Charles H and Bartolotta, John P and Bohn, Matthew and Chertkov, Eli and others},
  journal={Physical Review X},
  volume={15},
  number={2},
  pages={021052},
  year={2025},
  publisher={APS},
  url={https://doi.org/10.1103/PhysRevX.15.021052}
}

@article{gidney2021stim,
  title={Stim: a fast stabilizer circuit simulator},
  author={Gidney, Craig},
  journal={Quantum},
  volume={5},
  pages={497},
  year={2021},
  publisher={Verein zur F{\"o}rderung des Open Access Publizierens in den Quantenwissenschaften},
  url={https://doi.org/10.22331/q-2021-07-06-497}
}

@article{weinstein2012shor,
  title   = {On the use of {Shor} states for the [7,1,3] quantum error correcting code},
  author  = {Weinstein, Yaakov S. and Buchbinder, Sidney D.},
  journal = {Physical Review A},
  volume  = {86},
  number  = {5},
  pages   = {052336},
  year    = {2012},
  doi     = {10.1103/PhysRevA.86.052336},
  eprint  = {1111.3930},
  archivePrefix = {arXiv},
  primaryClass  = {quant-ph},
  url= {https://doi.org/10.1103/PhysRevA.86.052336}
}

@article{tomita2013comparison,
  title   = {Comparison of ancilla preparation and measurement procedures for the
             {Steane} [[7,1,3]] code on a model ion trap quantum computer},
  author  = {Tomita, Yu and Guti{\'e}rrez, Mauricio and Kabytayev, Chingiz and
             Brown, Kenneth R. and Hutsel, M. R. and Morris, A. P. and
             Stevens, Kelly E. and Mohler, G.},
  journal = {Physical Review A},
  volume  = {88},
  number  = {4},
  pages   = {042336},
  year    = {2013},
  doi     = {10.1103/PhysRevA.88.042336},
  eprint  = {1305.0349},
  archivePrefix = {arXiv},
  primaryClass  = {quant-ph},
  url = {https://doi.org/10.1103/PhysRevA.88.042336}
}

@article{calderbank1998quantum,
  title={{Quantum error correction via codes over GF (4)}},
  author={Calderbank, A Robert and Rains, Eric M and Shor, Peter M and Sloane, Neil JA},
  journal={IEEE Transactions on Information Theory},
  volume={44},
  number={4},
  pages={1369--1387},
  year={1998},
  publisher={IEEE},
  url={https://doi.org/10.1109/18.681315}
}

@phdthesis{gottesman1997stabilizer,
  title={Stabilizer codes and quantum error correction},
  author={Gottesman, Daniel},
  year={1997},
  type = {PhD dissertation},
  publisher={California Institute of Technology},
  url={https://doi.org/10.7907/rzr7-dt72}
}
\clearpage
\appendix
\onecolumngrid
\section*{Appendices}
\section{Fault-equivalence proofs of the rewrite rules}
\label{app:der_rewrite}
We prove the fault equivalence of our rewrite rules using the fault-equivalent ZX-calculus rules
\textsc{Elim}$_{\mathrm{fe}}$, \textsc{Fuse-1}$_{\mathrm{fe}}$, and \textsc{OCM}
from Ref.~\cite{rodatz2025fault}.
\subsection{Restricted commutation}
\includegraphics[]{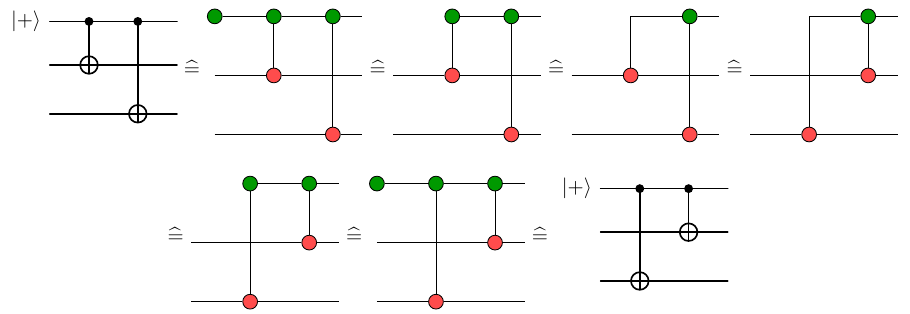}
\subsection{Basis swap}
\includegraphics[]{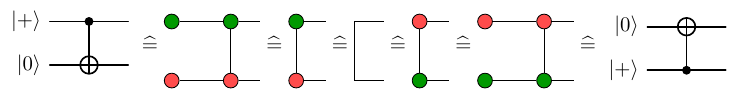}
\subsection{Target swap}
\includegraphics[]{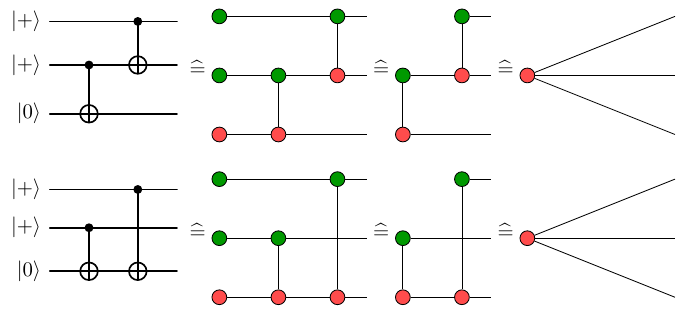}
\subsection{Bell reduction}
\includegraphics[]{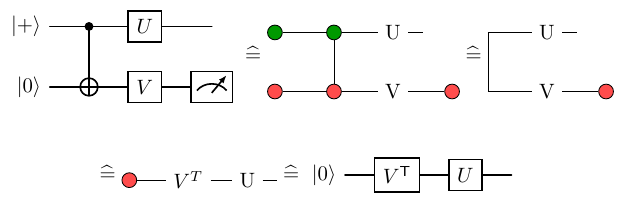}
\section{Derivation of optimized Shor gadget}
\label{app:shor}
\resizebox{\linewidth}{!}{
\includegraphics[]{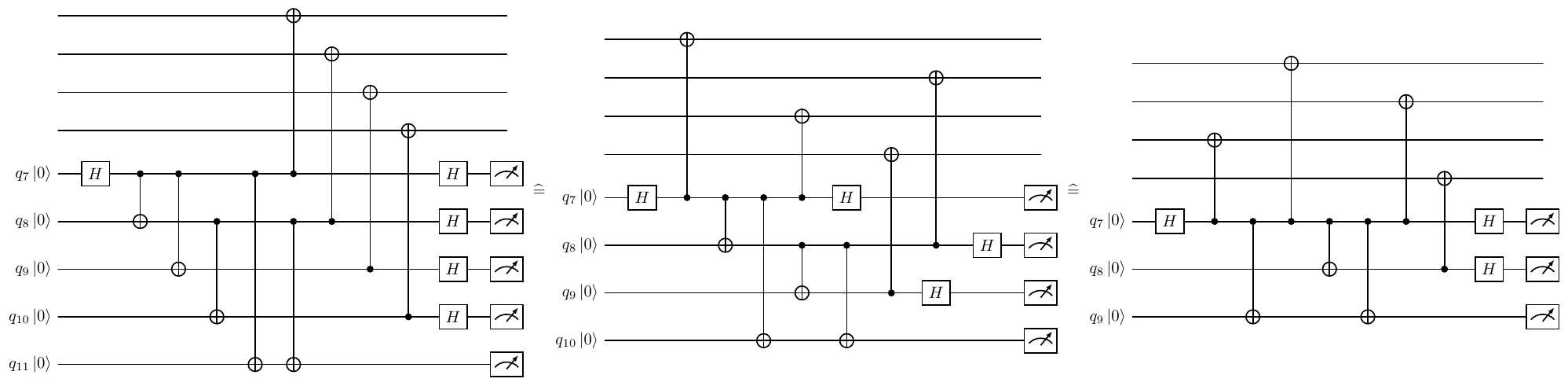}
}
Every gadget of the round reduces the same way, by the Bell reduction $(7,9)$ and $(7,8)$.
\section{Derivation of optimized Steane gadget}
\label{appendix:goto-derivation}
\resizebox{\linewidth}{!}{
\includegraphics[]{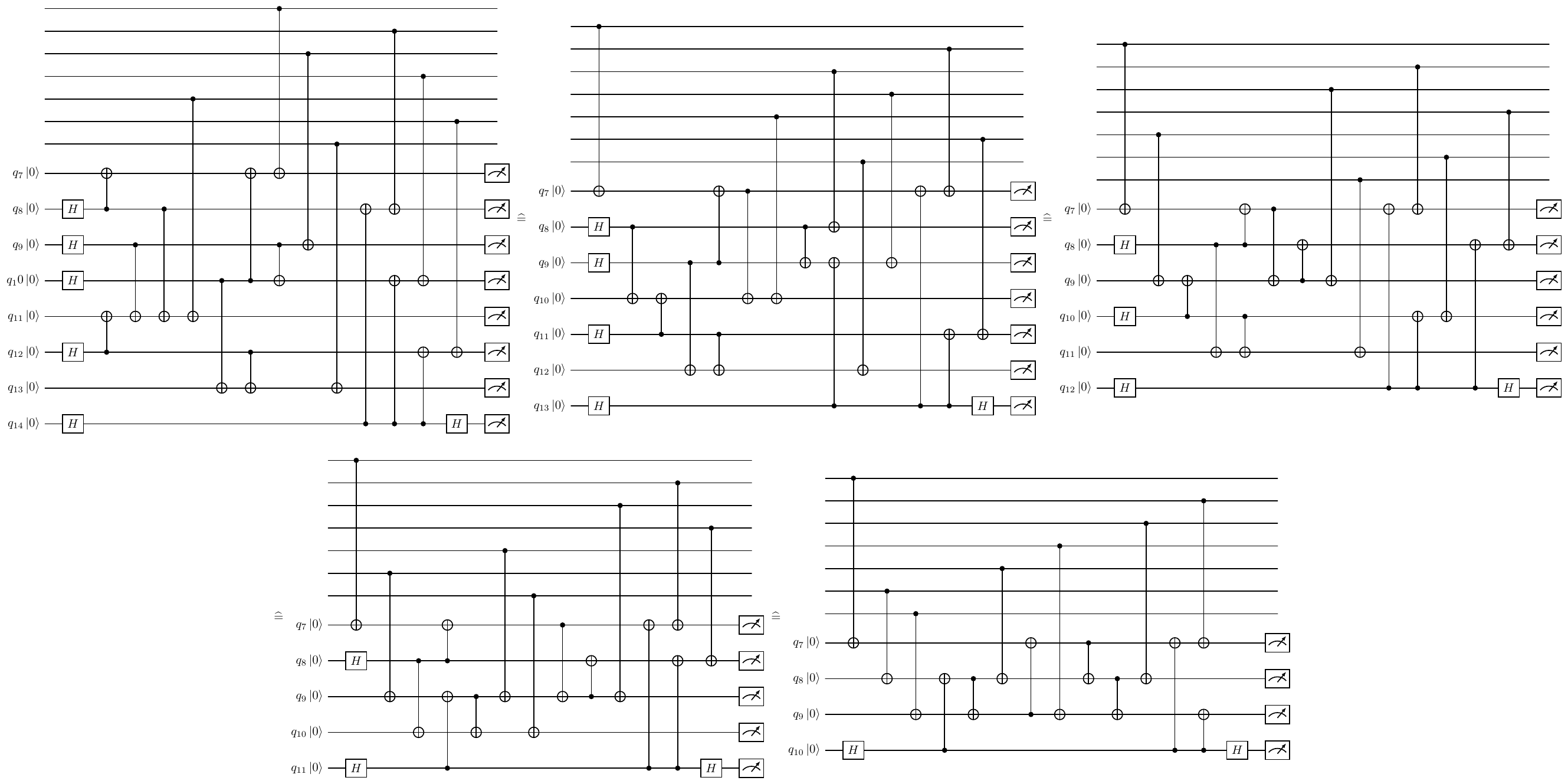}
}
We used four Bell reductions; $(7,8)$, $(10,8)$, $(10,9)$ and $(10,8)$.
\section{Derivation of the manually designed gadget}
\label{app:manual}
In this section, we reproduce the circuit of Po{\'o}r \textit{et al.}~\cite{poor2026ultra} using our rewrite rules. This derivation demonstrates that, despite being coarse-grained, our rules are sufficiently expressive to recover a circuit originally obtained through manual rewriting.
\begin{figure}
\includegraphics[width=\linewidth]{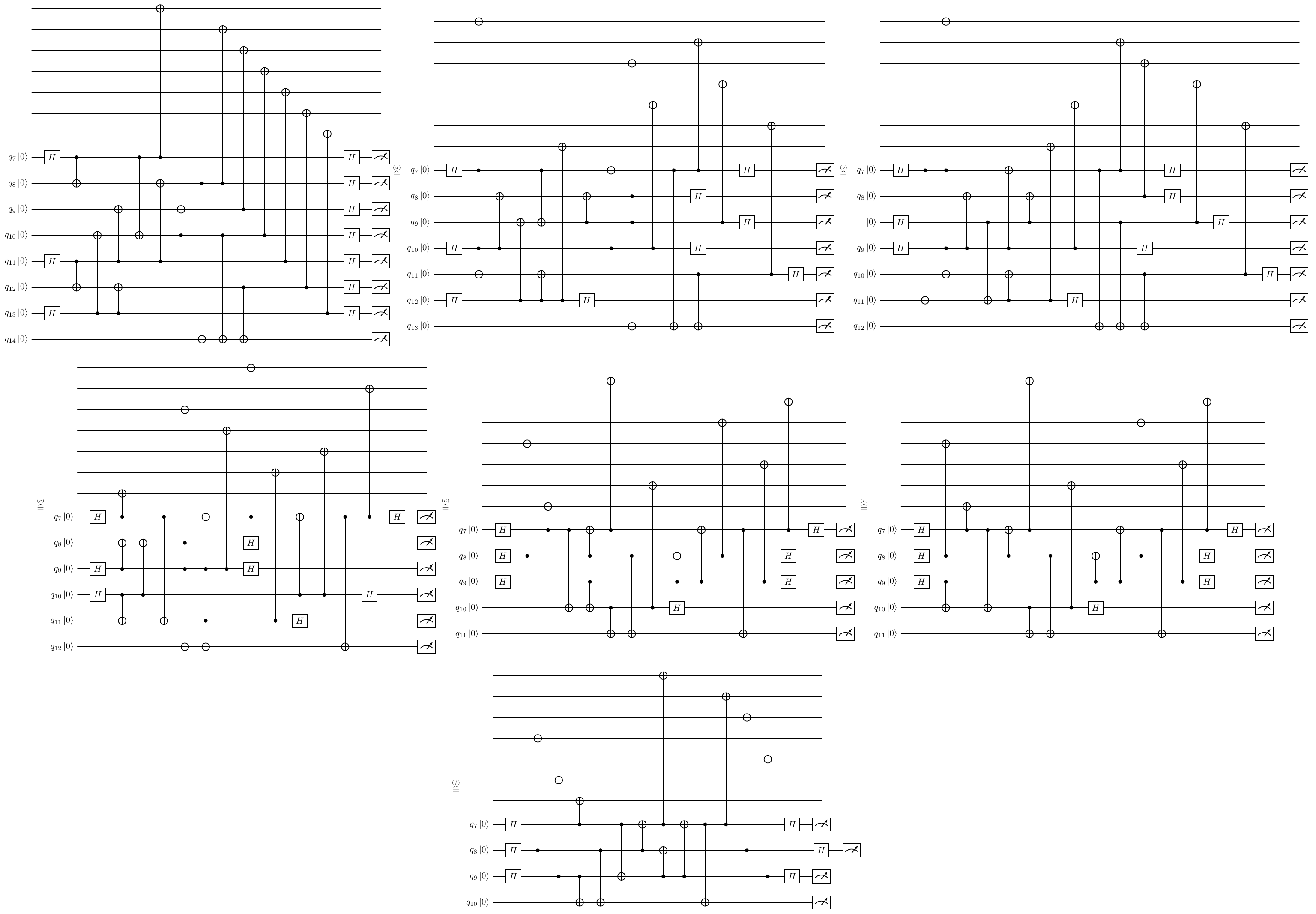}
\caption{Derivation of manually designed circuit}
\end{figure}
The steps in the diagram apply the following rewrites:
\begin{enumerate}
\item[(a)] Bell reduction on qubits $(7,8)$.
\item[(b)] Restricted commutation of \textsc{cnot}s controlled by qubit $7$, followed by a basis swap on qubits $(9,12)$ and a target swap of $\textsc{cnot}_{7\to9}$.
\item[(c)] Bell reduction on qubits $(7,12)$.
\item[(d)] Bell reduction on qubits $(9,8)$.
\item[(e)] Restricted commutation of \textsc{cnot}s targeting qubit $10$.
\item[(f)] Bell reduction on qubits $(10,11)$.
\end{enumerate}
\section{Performance across QEC cycle counts and physical error rates}
\label{app:data}
This appendix presents the parameter scans supporting the comparisons in Section~\ref{sec:evaluation}. We evaluate logical memories initialized in $\ket{0}_L$ and $\ket{+}_L$, varying either the physical two-qubit error rate $p$ at a fixed number of QEC cycles or the number of cycles $N$ at a fixed error rate. Other physical error rates scale with $p$ as specified in Section~\ref{sec:evaluation}.
\begin{figure}
    \centering
    \includegraphics[width=\linewidth]{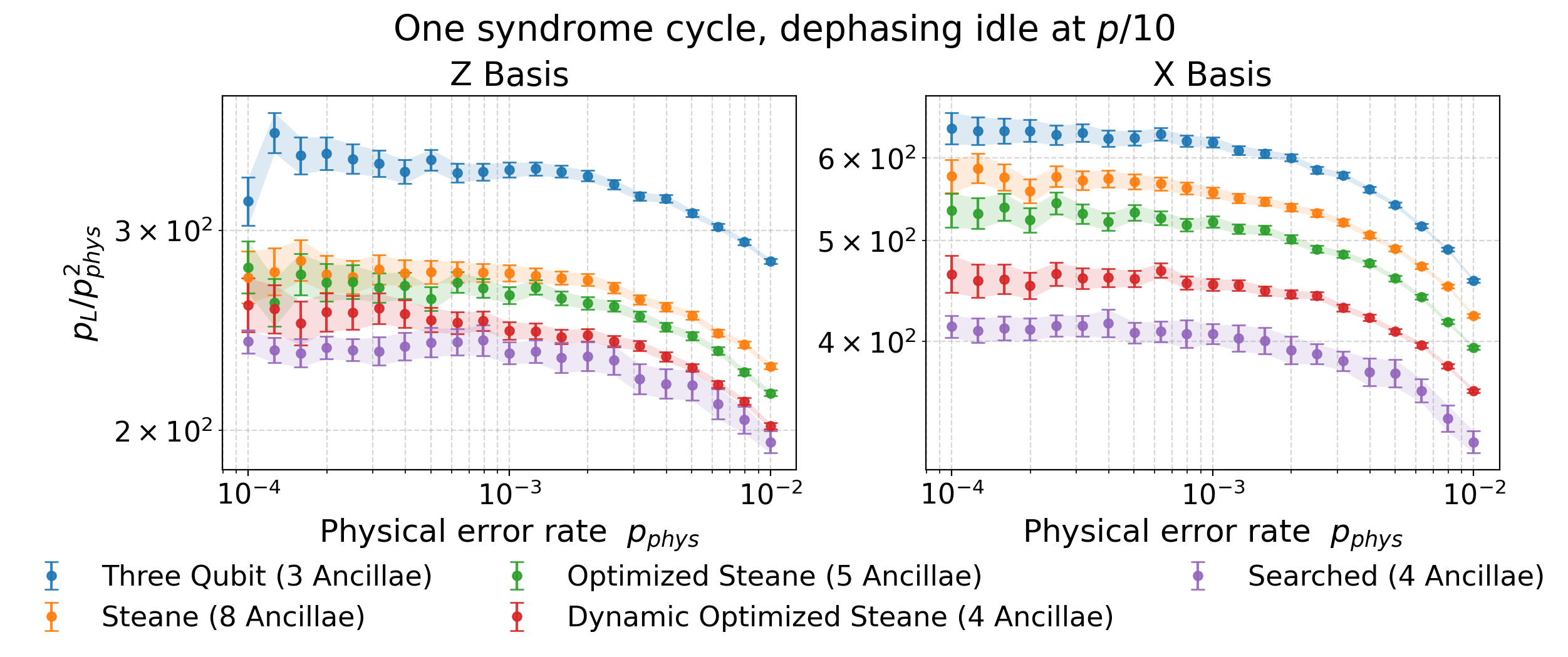}
    \caption{Logical error rates of five syndrome-extraction protocols for the $[[7,1,3]]$ code as a function of the physical two-qubit gate error rate $p$, for one QEC cycle ($N=1$). The comparison includes the three-qubit, Steane, optimized Steane, dynamic optimized Steane, and searched protocols. Idle qubits undergo dephasing noise with probability $p/10$. Results are shown for both $\ket{0}_L$ and $\ket{+}_L$ memories.}
\end{figure}
\begin{figure}
    \centering
    \includegraphics[width=\linewidth]{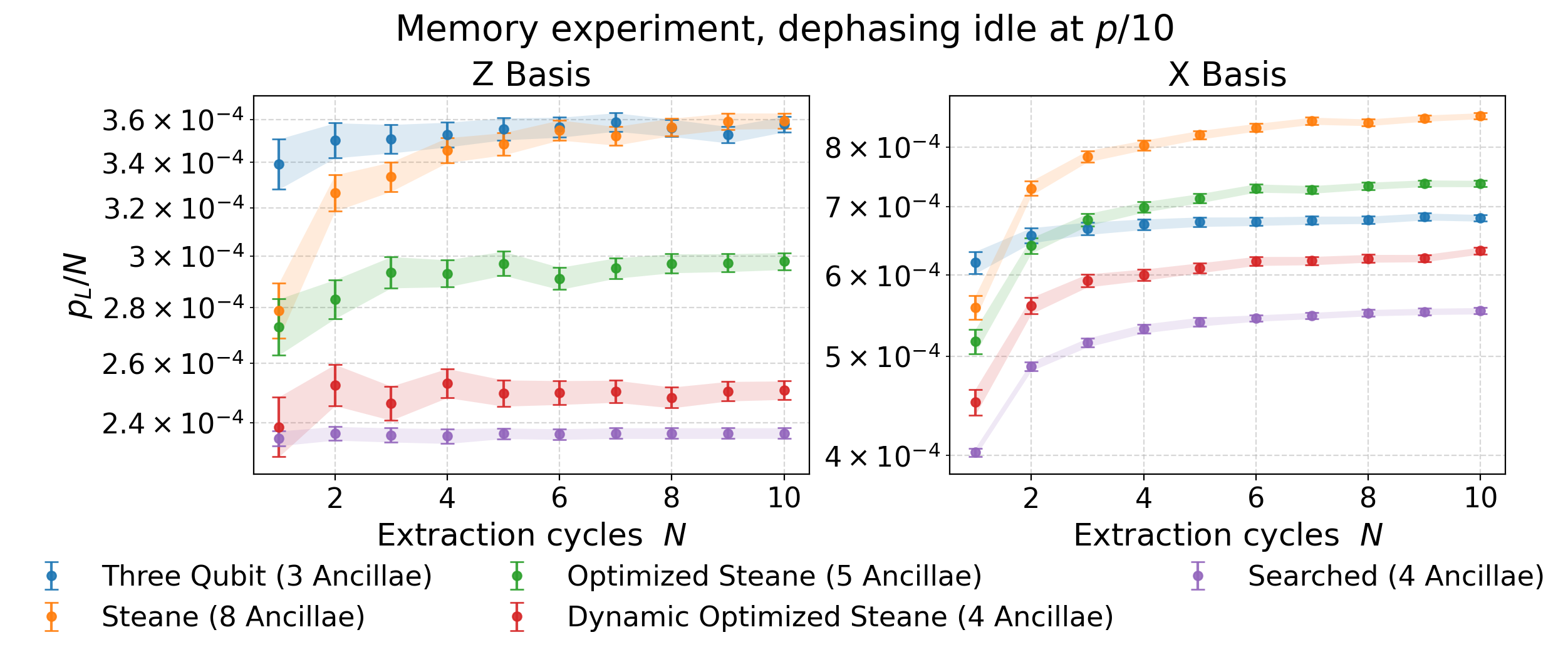}
    \caption{Logical error rates of the three-qubit, Steane, optimized Steane, dynamic optimized Steane, and searched protocols as a function of the number of QEC cycles, $N=1,\ldots,10$, at $p=10^{-3}$. Idle qubits undergo dephasing noise with probability $p/10$. Results for $\ket{0}_L$ and $\ket{+}_L$ memories show how logical errors accumulate under repeated syndrome extraction.}
\end{figure}
\begin{figure}
    \centering
    \includegraphics[width=\linewidth]{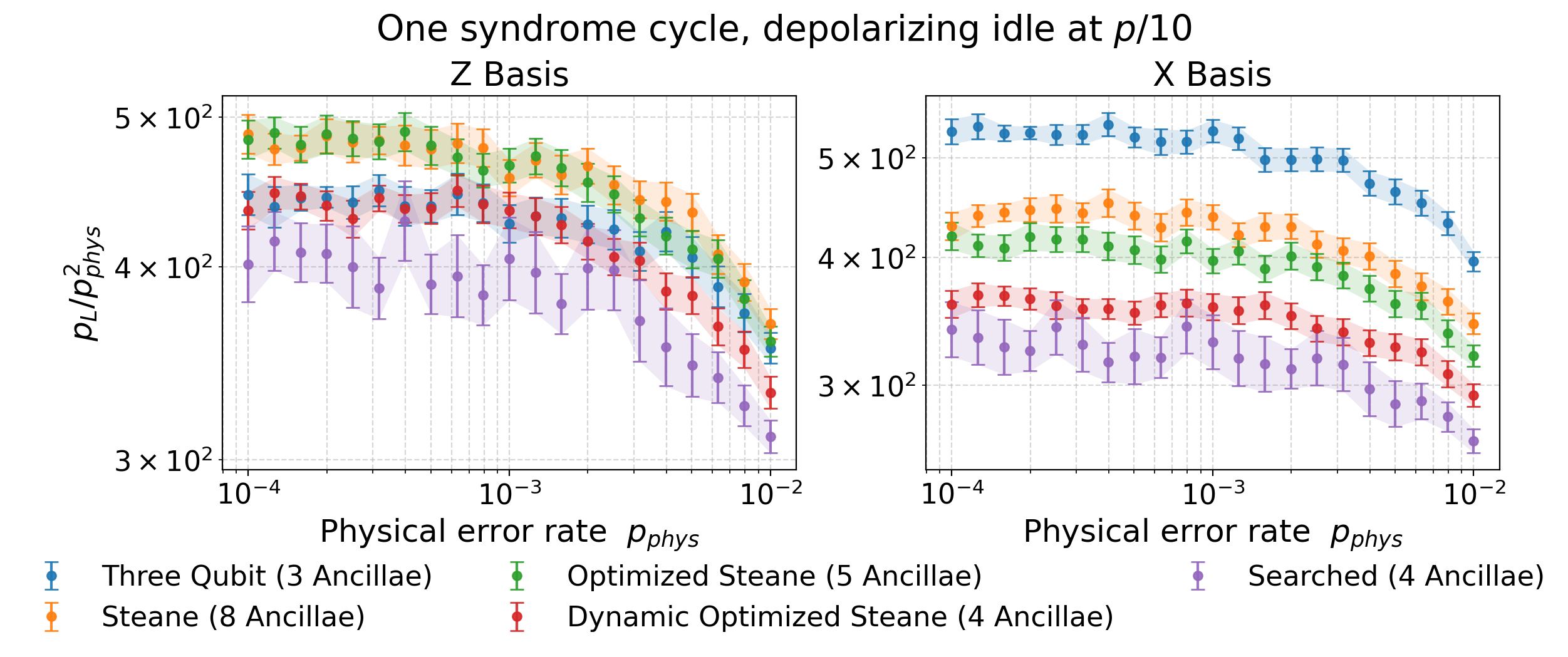}
    \caption{Logical error rates of the three-qubit, Steane, optimized Steane, dynamic optimized Steane, and searched protocols as a function of the physical two-qubit gate error rate $p$, for one QEC cycle ($N=1$). Idle qubits undergo depolarizing noise with total probability $p/10$, corresponding to probability $p/30$ for each of the Pauli errors $X$, $Y$, and $Z$. Results are shown for both $\ket{0}_L$ and $\ket{+}_L$ memories.}
\end{figure}
\begin{figure}
    \centering
    \includegraphics[width=\linewidth]{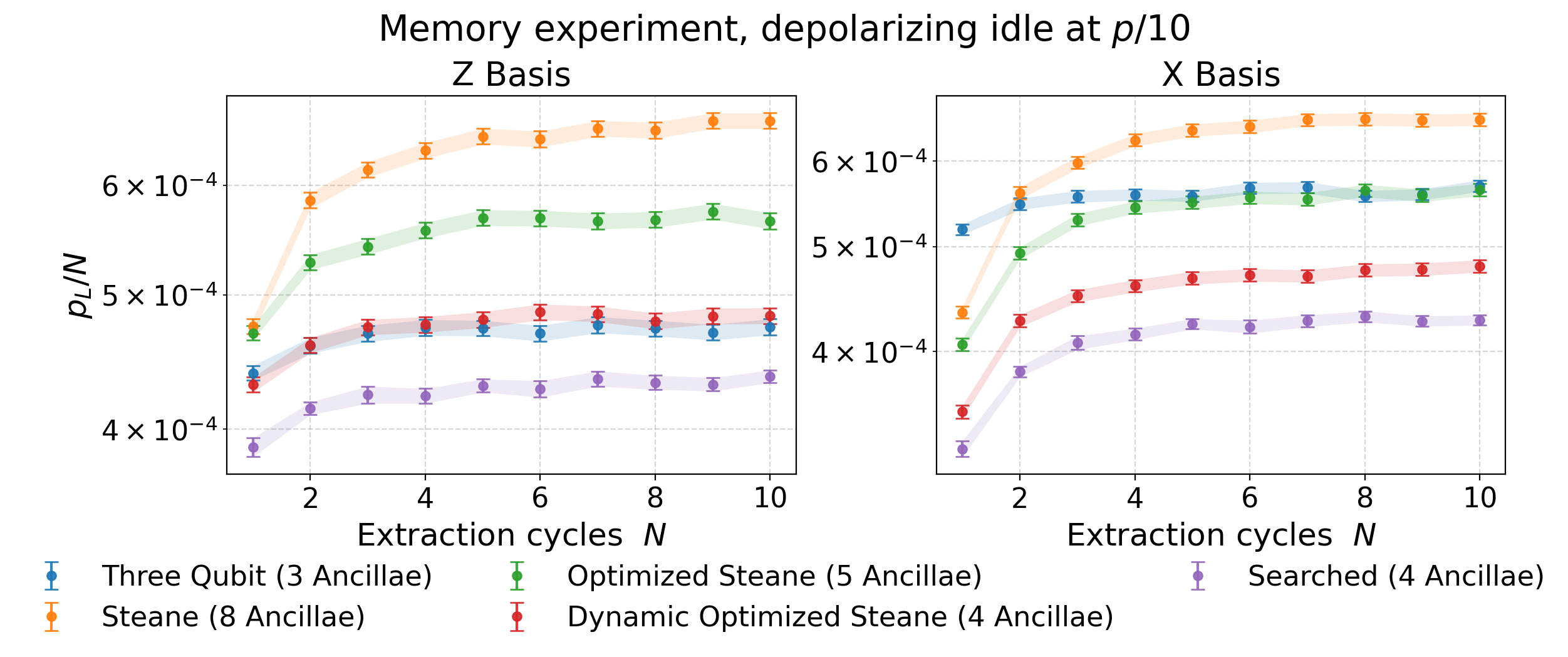}
    \caption{Logical error rates of the three-qubit, Steane, optimized Steane, dynamic optimized Steane, and searched protocols as a function of the number of QEC cycles, $N=1,\ldots,10$, at $p=10^{-3}$. Idle qubits undergo depolarizing noise with total probability $p/10$. Results are shown for $\ket{0}_L$ and $\ket{+}_L$ memories, allowing comparison with the dephasing model at the same total idle-error probability.}
\end{figure}
\begin{figure}
    \centering
    \includegraphics[width=\linewidth]{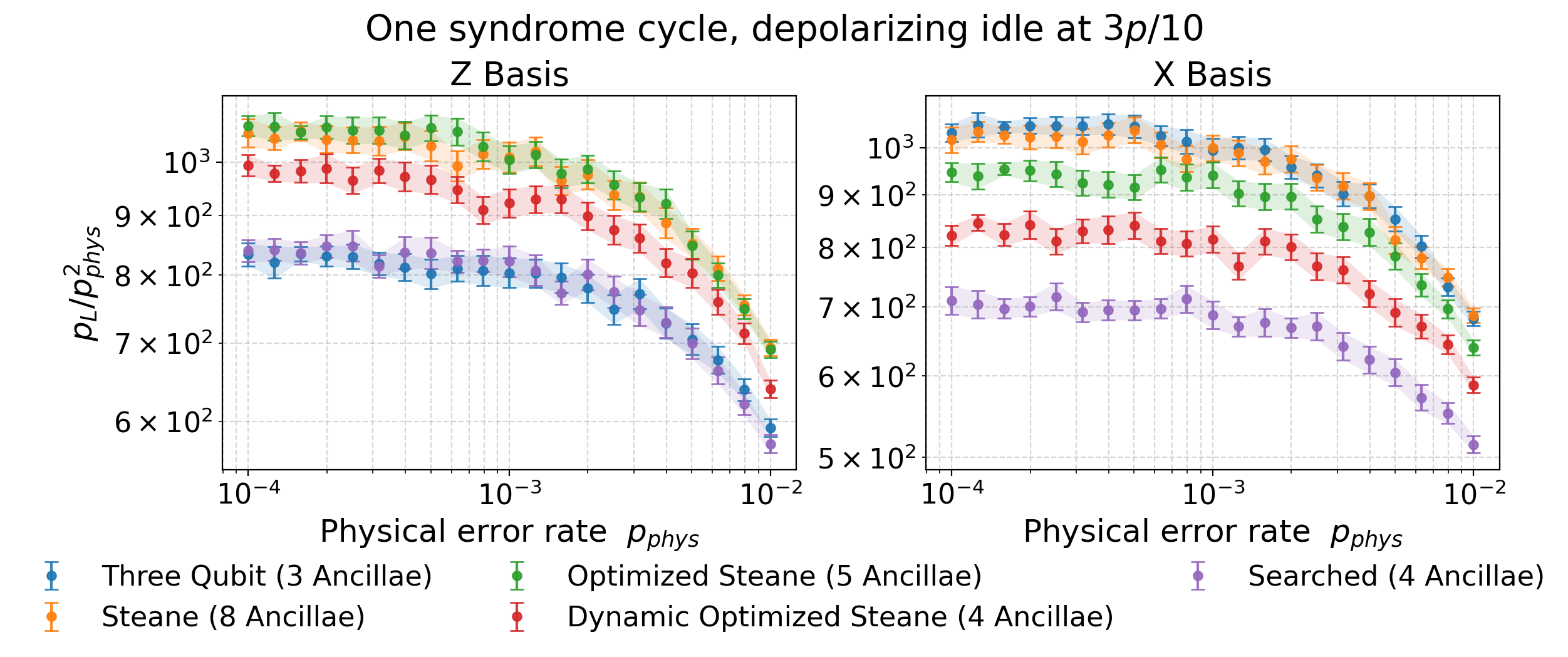}
    \caption{Logical error rates of the three-qubit, Steane, optimized Steane, dynamic optimized Steane, and searched protocols as a function of the physical two-qubit gate error rate $p$, for one QEC cycle ($N=1$). Idle qubits undergo depolarizing noise with total probability $3p/10$, corresponding to probability $p/10$ for each Pauli error. This matches the probability of a $Z$ error in the dephasing model while additionally introducing $X$ and $Y$ errors. Results are shown for both $\ket{0}_L$ and $\ket{+}_L$ memories.}
\end{figure}
\begin{figure}
    \centering
    \includegraphics[width=\linewidth]{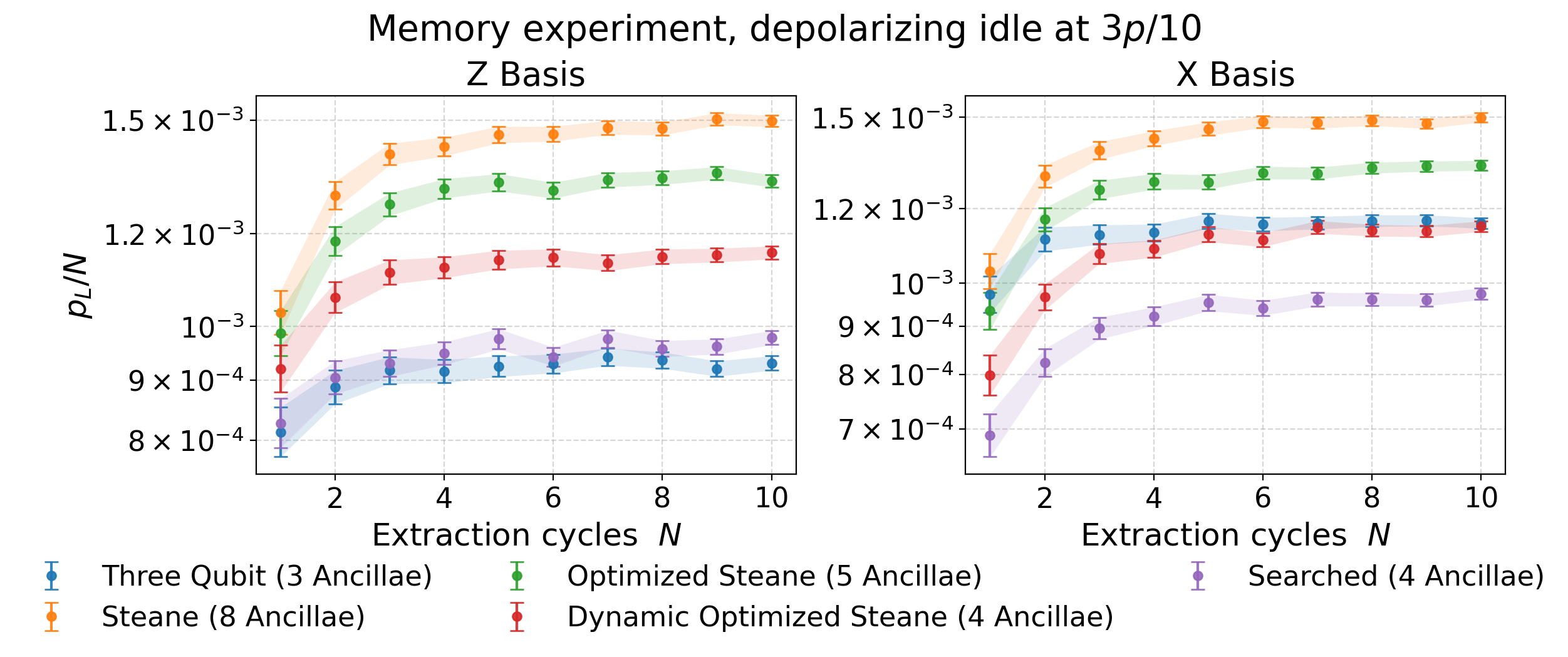}
    \caption{Logical error rates of the three-qubit, Steane, optimized Steane, dynamic optimized Steane, and searched protocols as a function of the number of QEC cycles, $N=1,\ldots,10$, at $p=10^{-3}$. Idle qubits undergo depolarizing noise with total probability $3p/10$, giving each Pauli error probability $p/10$. Results are shown for both $\ket{0}_L$ and $\ket{+}_L$ memories.}
\end{figure}
\end{document}